\pdfoutput=1

\documentclass[aip,amsmath,amssymb,reprint,nofootinbib,floatfix]{revtex4-1}

\usepackage{graphicx} % Include figure files
\usepackage{bm}% bold math
\usepackage[utf8]{inputenc}
\usepackage[T1]{fontenc}
\usepackage{mathptmx}
\usepackage{etoolbox}

\begin{document}

\title[Non-equilibrium classical recombination]%
{Non-equilibrium classical recombination in the expanding ultracold plasmas}

\author{Yurii V. Dumin}
\email[Electronic mail: ]{dumin@pks.mpg.de, dumin@yahoo.com}
\affiliation{Lomonosov Moscow State University,
Sternberg Astronomical Institute, \\
Universitetskii prosp.\ 13, 119234 Moscow, Russia}
\affiliation{Space Research Institute of Russian Academy of Sciences, \\
Profsoyuznaya str.\ 84/32, 117997 Moscow, Russia}
% \affiliation{Max Planck Institute for the Physics of Complex Systems, \\
% Noethnitzer Str.\ 38, 01187 Dresden, Germany}
\author{Ludmila M. Svirskaya}
\email[Electronic mail: ]{svirskaialm@susu.ru, svirskayalm@mail.ru}
\affiliation{South Ural State University,
Prosp.\ Lenina 76, 454080 Chelyabinsk, Russia}
\affiliation{South Ural State Humanitarian Pedagogical University, \\
Prosp.\ Lenina 69, 454080 Chelyabinsk, Russia}
\author{Eugen S. Savinykh}
\email[Electronic mail: ]{z-0000001@hotmail.com}
\affiliation{Retired}

\date{6 April 2026}

\begin{abstract}
The efficiency of recombination is of crucial importance for the existence
of ultracold plasmas (UCP), particularly, the ones formed in
the magneto--optical traps.
Unfortunately, the equilibrium thermodynamic treatment of
the ionization--recombination processes is inappropriate for the evolving
UCP clouds, while the straightforward kinetic simulation encounters
the problem of huge difference in the spatial and temporal scales for
free and bound motion of the electrons.
As a result, only the ``virtual'' electron--ion pairs are usually reproduced
in such modeling, and it is necessary to employ some heuristic criteria
to identify them with the recombined atoms.
It is the aim of this paper to present the first successful \textit{ab initio}
simulation of the non-equilibrium recombination in the evolving UCP plasmas.
We employed a special algorithm, which is based on using the ``scalable''
reference frame, co-moving with the expanding substance.
Then, the recombination events are identified by a series of sharp
equidistant peaks in the kinetic and/or potential energies, which are
caused by the captured electrons passing near the pericenters of their
orbits; and this is confirmed by a detailed inspection of their
trajectories.
Thereby, we were able to trace the real---rather than
``virtual''---electron--ion pairs, and the total efficiency of their
formation was found to be about~20\%, which is in agreement with
the laboratory measurements.
\end{abstract}

\maketitle

% ====================
\section{Introduction}
\label{sec:Intro}
% ====================

The ultracold plasmas (UCP), formed after release of the bunches of
photoionized atoms from the magneto--optical traps~\cite{Killian99},
represent a relatively new branch of plasma physics, emerged in the very
late 1990s and early 2000s~\cite{Gould01,Bergeson03,Killian07a,Killian07b}.
The possibility of existence of such plasmas for an appreciable time
interval crucially depends on the efficiency of recombination, since it
might be intuitively expected that the rarefied charged particles with
a sufficiently strong Coulomb coupling will quickly collapse into
the neutral pairs.

It is interesting to mention that the first molecular-dynamic simulations of
such plasmas started already in the early 1990s, \textit{i.e.}, a decade
before the first experiments with UCP.
Surprisingly, these simulations demonstrated that the ultracold plasmas
experienced a very small recombination, \textit{i.e.}, could survive for
a sufficiently long time~\cite{Mayorov94}.
Unfortunately, it was suspected by some opponents that the anomalous
resistance of charged particles with respect to the recombination might be
just an artifact of the particular numerical methods.
For example, the recombined neural pairs could be efficiently destroyed by
the collisions with walls of the computational box~\cite{Ignatov95,Mayorov95},
if the reflective boundary conditions were imposed there.
Moreover, the anomalously small recombination rate might be caused just by
the accumulation of numerical errors, preventing formation of the bound
electron--ion pairs.
This problem is especially severe because of the huge difference in
the characteristic temporal and spatial scales for the free and bound motion
of the charged particles in plasmas.

In view of the above-mentioned problem, the most of works on recombination
in UCP employed the artificial two-step procedure:
Firstly, the electron distribution function was simulated numerically by
the molecular-dynamic approach, and then it was used for analytical
estimates of the recombination rate, \textit{e.g.}, employing the idea of
electron diffusion in the energetic space~\cite{Bobrov11}.

On the other hand, as regards \textit{ab initio} modeling of recombination,
the most of works employed some additional criteria for the formation of
the bound electron--ion pairs:
For example, it was assumed that the act of recombination took place if
an electron performed a certain number of revolutions about the nearest
ion (\textit{e.g.}, 4~complete revolutions)~\cite{Lankin09b}.
Yet another approach was to assume that all electrons located closer
than a specified distance from the nearest ions (\textit{e.g.}, 20\% of
the Wigner--Seitz radius) will ultimately recombine~\cite{Niffenegger11}.
These artificial criteria were necessary because the commonly-used numerical
schemes did not allow ones to get sufficiently stable electron--ion pairs:
they were destroyed rather quickly due to accumulation of the computational
errors.
Besides, these works usually performed the simulations within a box of
a fixed size, which is poorly relevant to the actual experimental conditions,
since the UCP clouds typically expand by one or two orders of magnitude in
the course of experiments.

It is the aim of the present work to employ a specially designed numerical
scheme that is able, firstly, to take into account the considerable degree of
expansion of the plasma cloud and, secondly, to trace formation of the stable
electron--ion pairs (which should be subsequently converted to neutral atoms
due to the quantum processes).
In fact, the basic idea of this algorithm was already used in our earlier
paper~\cite{Dumin11}, which was aimed at \textit{ab initio} simulation of
the electron temperature evolution in UCP bunches released from
the magneto--optical traps.
The corresponding functional dependence obtained in that work,
$ T_e \propto t^{\alpha} $ with exponent $ \alpha \in [-1.25, -1.08] $,
was in perfect agreement with the experiment~\cite{Fletcher07}, where
$ \alpha = -(1.2{\pm}0.1) $.
It is important to emphasize that this coincidence was achieved exactly from
``the first principles'', without postulation of any additional heat sources
or sinks.
As will be shown below, the same numerical scheme provides very good results
also in the simulation of recombination.

% ====================
\section{The algorithm of simulation}
\label{sec:Algorithm}
% ====================

% ====================
\subsection{Geometrical configuration}
\label{sec:Geometry}
% ====================

%%%%%%%%%%%%%%%%%%%%%%%%%%%%%%%%%%%%%
\begin{figure}
\includegraphics[width=1.0\columnwidth]{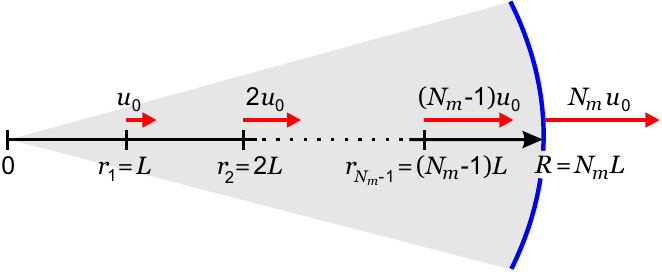}
\caption{\label{fig:Cloud}
Sketch of the plasma cloud (gray segment with a blue boundary) composed of
the identical cells of size~$ L $, uniformly expanding with
velocity~$ u_0 $.}
\end{figure}
%%%%%%%%%%%%%%%%%%%%%%%%%%%%%%%%%%%%%

Following the design of typical UCP experiments, we shall consider a large
volume of plasma expanding uniformly in all directions, starting from
the center $ r = 0 $.
For the sake of simplicity, we assume that the cloud is uniform and has
a sharp boundary at $ r = R(t) $; see Fig.~\ref{fig:Cloud}.
Next, let this cloud be composed of the identical (or ``mirror'') cubic cells
filled with some number of ions~$ N $ and the same number of electrons,
and $ N_{\rm m} $~be the number of cells from the center to the boundary.
Each cell expands with velocity~$ u_0 $ with respect to its opposite face.
So, at the cloud boundary $ R = N_{\rm m} L $, the velocity becomes
$ V = N_{\rm m} \, u_0 $.
In fact, this scheme closely resembles the idea of Hubble--Lema{\^i}tre
expansion, commonly used in cosmology~\cite{Zeldovich83}.

It is well known that the initial thermal energy of plasma (which, in
the case of photoionization, is concentrated mostly in electrons) quickly
transforms into kinetic energy of macroscopic motion of the cloud;
\textit{e.g.}, papers~\cite{Vikhrov20,Vikhrov21} and references therein.
In other words,
$ m_i \, V^2 \sim \, k_{\rm B} \, T_{e0} $,
where
$ T_{e0} $~is the initial electron temperature,
$ V $~is the velocity of the cloud boundary,
$ m_i $~is the ionic mass, and
$ k_{\rm B} $~is Boltzmann constant.
Consequently,
\begin{equation}
V \sim \sqrt{k_{\rm B} \, T_{e0} / m_i}
\end{equation}
(we did not write here any numerical coefficient, since it depends
substantially on the distribution of density inside a more realistic
cloud).
Just this formula can be used to link the experimental quantities
with the simulation parameters.

% ====================
\subsection{Equations of motion}
\label{sec:Eq_motion}
% ====================

%%%%%%%%%%%%%%%%%%%%%%%%%%%%%%%%%%%%%
\begin{figure}
\includegraphics[width=0.8\columnwidth]{Dumin_fig2.pdf}
\caption{\label{fig:Geom_config}
Sketch of calculation of the Coulomb sums, where each electron~$ e $
interacts not only with a specified ion~$ i $ inside the basic cell
(solid arrow) but also with infinite number of its ``mirror''
images~$ i^\prime $ (dashed arrows).}
\end{figure}
%%%%%%%%%%%%%%%%%%%%%%%%%%%%%%%%%%%%%

The simulations are based on solving the classical equations of
motion for $ N $~electrons inside the ``basic'' cell:
\begin{equation}
m_e \, \frac{d^2 {\rm \bf r}_i}{d t^2} = \, {\rm \bf F}_i \, ,
\label{eq:el_motion_orig}
\end{equation}
where $ {\rm \bf r}_i $~are the electron coordinates, and
$ m_e $~is the electron mass.
The Coulomb force~$ {\rm \bf F}_i $, acting on the $i$'th electron,
is calculated taking into account all particles in the system
\textit{i.e.}, not only within the same basic cell but also in all its
``mirror'' images (Fig.~\ref{fig:Geom_config}):
\begin{eqnarray}
{\bf F}_i & = &
  \sum_{j=1}^{N} e^2 \,
    \frac{{\bf R}_j - {\bf r}_i}{|{\bf R}_j - {\bf r}_i|^3} \; + \,
  \sum_{j=1, \, j \neq i}^{N} e^2 \,
    \frac{{\bf r}_i - {\bf r}_j}{|{\bf r}_i - {\bf r}_j|^3}
\nonumber \\
& + &
\sum_{{\bf n} \, \in \, \mathbb{Z}^3 \backslash \{ {\bf 0} \}}
  \Bigg\{ 
    \sum_{j=1}^{N}  e^2
      \frac{( {\bf R}_j + L \, {\bf n} ) - {\bf r}_i )}%
      {\big| ( {\bf R}_j + L \, {\bf n} ) - {\bf r}_i\big|^3}
\nonumber \\
& &
\;\;\;\;\;\;\;\;\;\;\;\;\;\;\:
  + \; \sum_{j=1}^{N} \, e^2
    \frac{{\bf r}_i - ( {\bf r}_j + L \, {\bf n} )}%
    {\big|{\bf r}_i - ( {\bf r}_j + L \, {\bf n} )\big|^3}
    \Bigg\} \: .
\label{eq:electron_forces}
\end{eqnarray}
Here,
$ {\bf R}_j \; (j = 1,\dots,N) $~are the ionic coordinates,
$ N $~is the number of charged particles of each kind in the basic cell,
$ e $~is the electron charge,
$ L $~is the linear size of the basic cell, and
$ \mathbb{Z}^3 \backslash \{ {\bf 0} \} $~denotes the three-dimensional
array of integers without the zero element.
We do not write here a similar formula for the forces acting on ions:
since the ions are much more massive than electrons, they can be usually
assumed to move by inertia.

%%%%%%%%%%%%%%%%%%%%%%%%%%%%%%%%%%%%%
\begin{figure}
\includegraphics[width=1.0\columnwidth]{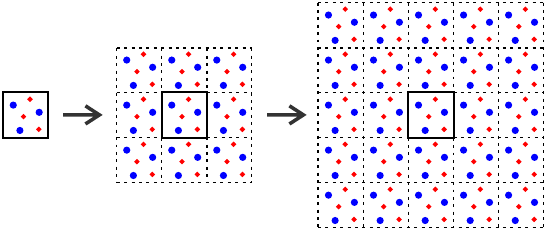}
\caption{\label{fig:Sum_mir_cell}
Sketch of summation over the mirror cells (dotted squares) around
the basic cell (solid square) in calculation of the Coulomb forces.}
\end{figure}
%%%%%%%%%%%%%%%%%%%%%%%%%%%%%%%%%%%%%

In fact, summation in formula~(\ref{eq:electron_forces}) is performed over
a finite number of mirror cells until the specified accuracy of convergence
of the Coulomb forces is achieved.
These mirror cells are arranged into the cubic shells of increasing radii
around the basic cell, as illustrated in Fig.~\ref{fig:Sum_mir_cell}.
Strictly speaking, the sums like~(\ref{eq:electron_forces}) are conditionally
convergent, \textit{i.e.}, the result can depend on the summation order.
The method of summation depicted in the figure seems most reasonable from
the physical point of view.
Really, effect of the mirror cells located at the larger radii should be
obscured by some additional physical perturbations, thereby providing
a ``physical regularization'' of the conditionally-convergent Coulomb sums.

Although we integrated the equations of motion for a relatively small number
of charged particles in the basic cell (\textit{e.g.}, a few dozens or a few
hundreds), the total number of particles taken into account in calculation
of the Coulomb sums was huge.
For example, if the relative accuracy of calculations was specified to be
about~$ 10^{-4} $ or~$ 10^{-5} $, then even at $ N = 10 $ the total number
of particles involved in the computation turned out to be from a few hundred
thousand to a few million.
Let us emphasize that this number was not fixed but chosen ``dynamically'' at
each step of integration.

In fact, the same type of summation as in formula~(\ref{eq:electron_forces})
is commonly implemented in the condensed-matter physics by the so-called
Ewald method.
(Its original idea is outlined, \textit{e.g.}, in textbook~\cite{Ziman72},
and a more elaborated version can be found in paper~\cite{Demyanov22}.)
Most probably, the same approach can be applied also to the problem under
consideration, but this issue requires a further study.

% ====================
\subsection{The scalable reference frame}
\label{sec:Scalable_frames}
% ====================

Yet another serious problem in simulation of the expanding plasma clouds is
a considerable change in the spatial scale of the system---and, therefore, in
the Coulomb forces---in the course of time, which makes it difficult to keep
the accuracy of integration at a stable level.
To get around this obstacle, we introduce the scalable reference frame,
expanding with the average expansion rate of the plasma.
Namely, a length of the basic cell is assumed to increase as
\begin{equation}
L(t) = L_0 + u_0 \, t ,
\label{eq:basic_cell_expansion}
\end{equation}
where
$ L_0 $~is its initial value, and
$ u_0 $~is the velocity of expansion.
From the physical point of view, such a linear law corresponds to the inertial
stage of plasma motion, when the most part of its initial thermal energy
was transformed into kinetic energy of the macroscopic expansion.
In fact, this law is established very quickly, and it is commonly used in
the interpretation of all experiments with UCP.

Besides, it is convenient to introduce the following dimensionless variables.
Let the unit of length be the time-dependent characteristic interparticle
separation:
\begin{equation}
\tilde{l}(t) = L(t) / {(2N)^{1/3}} \, .
\label{eq:length_unit}
\end{equation}
The unit of time is taken to be proportional to the characteristic period of
oscillations of an electron with mass~$ m_e $ in the field of the nearest
ion at the initial instant of time:
\begin{equation}
\tau = \big( m_e^{1/2} \! / e \big) \, {\tilde{l_0}}^{\, 3/2}
\label{eq:time_unit}
\end{equation}
(as distinct from the unit of length, it is independent of time).
From here on, all physical quantities normalized to~$ \tilde{l} $ and
$ \tau $ will be marked by asterisks.

As can be easily seen, the unit of length varies with the dimensionless
time as
\begin{equation}
\tilde{l} = \: \tilde{l_0} \, ( 1 + \, u_0^* \, t^* ) \, ,
\label{eq:length_unit_time}
\end{equation}
where
\begin{equation}
u_0^* = \, u_0 \, \tau / L_0
\label{eq:expan_veloc_dimles}
\end{equation}
is the dimensionless velocity of expansion;
while the dimensionless size of the basic cell remains constant:
\begin{equation}
L^* = (2N)^{1/3} .
\label{eq:dimles_L}
\end{equation}
In other words, the simulated particles perform their motion in a box of
the fixed size~(\ref{eq:dimles_L}), thereby mitigating the problem of large
variation of the spatial coordinates in the case of considerable plasma
expansion.

As usual, the periodic boundary conditions are imposed at the box boundaries:
when one of the particles leaves the box through one of its faces,
the identical particle with the same velocity enters this box through
the opposite face.
(Let us emphasize that just the periodic boundary conditions are especially
important in the simulation of recombination, because the reflective
boundaries will lead to destruction of the electron--ion pairs that have
been already formed.)

The equation of motion of the $ i $'th electron~(\ref{eq:el_motion_orig})
in the scale-dependent dimensionless variables is reduced to
\begin{equation}
{\ddot{\rm \bf r}}_i^*
+ \, 2 \, u_0^* \, ( 1 + \, u^*_0 \, t^* )^{-1} \,
  {\dot{\rm \bf r}}_i^* = \,
( 1 + \, u^*_0 \, t^* )^{-3} \, {\rm \bf F}^*_i \, ,
\label{eq:el_motion_norm}
\end{equation}
where dot denotes a derivative with respect to~$ t^* $.
Therefore, expansion of the coordinate frame results in the appearance
of an effective viscous force, proportional to the electron velocity,
which is presented by the second term in the left-hand side of this
equation.
In fact, just this dissipative force is responsible for the electron
deceleration and its subsequent trapping by the nearest ion.

The Coulomb forces, originally given by formula~(\ref{eq:electron_forces}),
after the normalization are reduced to
\begin{eqnarray}
{\bf F}^*_i & = &
  \sum_{j=1}^{N} \,
    \frac{{\bf R}^*_j - {\bf r}^*_i}{|{\bf R}^*_j - {\bf r}^*_i|^3} \; + \,
  \sum_{j=1, \, j \neq i}^{N} \,
    \frac{{\bf r}^*_i - {\bf r}^*_j}{|{\bf r}^*_i - {\bf r}^*_j|^3}
\nonumber \\
& + &
\sum_{{\bf n} \, \in \, \mathbb{Z}^3 \backslash \{ {\bf 0} \}}
  \Bigg\{
    \sum_{j=1}^{N}
      \frac{( {\bf R}^*_j + L^* \, {\bf n} ) - {\bf r}^*_i )}%
      {\big| ( {\bf R}^*_j + L^* \, {\bf n} ) - {\bf r}^*_i\big|^3}
\nonumber \\
& &
\;\;\;\;\;\;\;\;\;\;\;\;\;\;\:
  + \; \sum_{j=1}^{N}
    \frac{{\bf r}^*_i - ( {\bf r}^*_j + L^* \, {\bf n} )}%
    {\big|{\bf r}^*_i - ( {\bf r}^*_j + L^* \, {\bf n} )\big|^3}
    \Bigg\} \: .
\label{eq:electron_forces_dimless}
\end{eqnarray}
As was already explained above, the sums over~$ \bf n $ are actually
calculated over the cubic shells of increasing radii, composed of
the ``mirror'' cells, until the required accuracy of convergence is
achieved.

% ====================
\subsection{The method of integration}
\label{sec:Integr_method}
% ====================

Following the standard procedure, the second-order differential
equation~(\ref{eq:el_motion_norm}) can be rewritten as a system of two
first-order differential equations:
\begin{subequations}
\begin{eqnarray}
{\dot{\rm \bf r}}_i^* & \!\!\!\! = & \!\!\!\! {\rm \bf v}_i^* ,
\label{eq:diff_eq1}
\\
{\dot{\rm \bf v}}_i^* & \!\!\!\! = &
  \!\!\!\! - 2 \, u_0^* s \, {\rm \bf v}_i^* + \,
  s^3 \, {\rm \bf F}^*_i ,
\label{eq:diff_eq2}
\end{eqnarray}
\end{subequations}
where
\begin{equation}
s(t^*) = \, ( 1 + \, u^*_0 \, t^* )^{-1} .
\label{eq:scale_fac}
\end{equation}

Next, they are integrated numerically by the second-order Runge--Kutta
method with a constant stepsize.
In fact, this method is rather inefficient, and a much better computational
performance could be achieved by using the higher-order integrators with
the adaptive stepsize control.
This was done, for example, in our previous works~\cite{Dumin22,Dumin24},
where we employed subroutines from the \textit{Numerical Recipes}
library~\cite{Press92}.
Unfortunately, the adaptive stepsize control operates reliably only for
the reflective boundaries and encounters serious problems in the case of
periodic boundary conditions (for more details, see Appendix~A in
paper~\cite{Dumin22}).
So, since the geometrical setup of our simulations (Fig.~\ref{fig:Geom_config})
inherently incorporates the periodic boundaries, we were enforced to use here
a rather primitive integration scheme.

Yet another very subtle point in our modeling is usage of the exact
(singular) Coulomb potentials of the charged particles, without any
``softening'' or cut-offs at the small distances, which was a common practice
in many earlier papers~\cite{Lankin09a,Niffenegger11,Tiwari17}.
We preferred to avoid such a simplification, because formation of the bound
states (resulting in the recombination) is rather sensitive to the specific
form of the potential.
Particularly, the closed trajectories are possible only in the exact Coulomb
field~\cite{Landau76}.
(Yet another option for the formation of closed orbits is the potential of
harmonic oscillator, proportional to~$ r^2 $; but this is evidently beyond
our physical context.)

Unfortunately, if the simulated plasma was initially cold enough, then
the electrons began to fall onto the nearest ions, sometimes resulting in
the almost head-on collisions, which could not be resolved by our integration
algorithm.
Such cases were identified by the sharp jumps in the kinetic energy of
electrons and their asymmetric temporal profiles (with respect to
the passage of pericenters); for the particular examples, see
Appendix~\ref{sec:Examples} in the present paper.
Then, the integration was cancelled, and the corresponding initial conditions
were discarded.
So, we used only the ``favorable'' initial conditions, which did not lead to
the failures.
In fact, the number of failures was a few times greater than the number of
successful attempts.
Fortunately, this did not increase appreciably the computational cost of
our simulations, because the failures typically occurred at the very early
stage of integration (about $ 10\,\tau $ from the beginning), while we were
interested in the long-term dynamics (several tens or hundreds~$ \tau $).
So, a few failures per one successful attempt did not take too much time.

% ====================
\subsection{Initial conditions}
\label{sec:Init_cond}
% ====================

An accurate specification of the initial conditions for ultracold plasmas is
a rather nontrivial tack.
For example, if the plasma is produced by \textit{instantaneous}
photoionization of neutral atoms, then the initial electron positions should
be substantially correlated with the ionic ones, and distribution of
the electron velocities will be non-Gaussian~\cite{Niffenegger11}.
In the opposite case, when the ionization process takes some time,
the released electrons will be mixed in space between the ions and
approximately thermalized, resulting in the uniform coordinate and Gaussian
velocity distributions.
Since we are interested in the recombination events occurring after a quite
long temporal evolution, exact specification of the initial conditions should
not be important.
So, for the sake of simplicity, we followed the second option.
Namely, initial positions of the electrons~$ r^*_{i0} $ and ions~$ R^*_{i0} $
($ i = 1,\dots, N $) were given by the random numbers distributed uniformly
within the box, $ -L^*/2 \le (x^*, y^*, z^*) \le L^*/2 $.

On the other hand, the initial electron velocities~$ {\dot r}^*_{i0} $ were
specified by the normal (Gaussian) distribution with a root-mean-square
deviation~$ {\sigma}_{v0}^* $ over each Cartesian coordinate.
This quantity evidently characterizes the initial electron temperature.
As was shown in our work~\cite{Dumin11}, the formal definition
$ T_e = (2/3)\,E_{\rm kin} / k_{\rm B} $
(where $ E_{\rm kin} $~is the electron kinetic energy, and $ k_{\rm B} $~is
Boltzmann constant) should work rather well even if the electrons strongly
interact with ions.
At last, as was already mentioned above, the ions are assumed to move by
inertia due to their very large masses, \textit{i.e.}, in the co-moving
reference frame they are at rest:
$ R^*_{i} \equiv \mbox{const} $ and~$ {\dot R}^*_{i} \equiv 0 $.

% ====================
\section{Results of the simulations}
\label{sec:Results}
% ====================

%%%%%%%%%%%%%%%%%%%%%%%%%%%%%%%%%%%%%
\begin{figure}
\includegraphics[width=0.9\columnwidth]{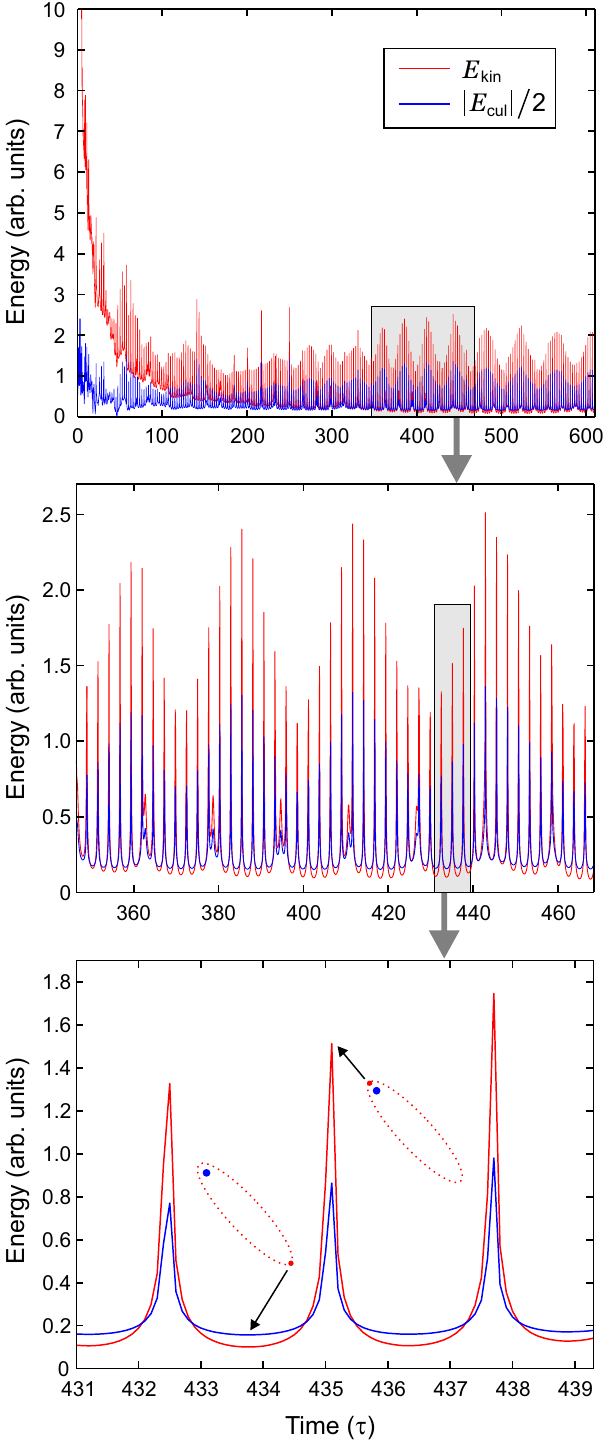}
\caption{\label{fig:Energy_time}
Temporal dependence of the kinetic (red curves) and potential (blue curves)
energies of all particles in the basic cell at various ``magnifications''
(successive panels, from top to bottom).
The regions of subsequent magnification are marked by the gray rectangles.}
\end{figure}
%%%%%%%%%%%%%%%%%%%%%%%%%%%%%%%%%%%%%

The simulations were performed for various numbers of the charged particles
of each kind~$ N $ in the basic cell, from 10 to 1000.
Since the employed algorithm of numerical integration without the adaptive
stepsize control was computationally expensive, a sufficiently long interval
of integration (about~$ 1000\,\tau $) was achieved only for the small number
of particles (\textit{e.g.}, $ N = 10 $); and the most convincing patterns of
recombination were obtained just in this case.
On the other hand, when $ N $ was substantially larger, the achievable
interval became shorter by an order of magnitude; so that only the initial
stage of recombination was observed, and it was difficult to derive a reliable
conclusion on its total efficiency.
(To avoid misunderstanding, let us remind once again that the actual number of
particles taken into account in calculation of the Coulomb forces, including
the ``mirror'' cells, was much greater: for example, even at $ N = 10 $ it
varied from a few hundred thousand to a few million, depending on the rate of
convergence of the Coulomb sums at each particular step of integration.)

In the results presented below, we used the initial dispersion of the electron
velocities $ {\sigma}_{v0}^* = 3.0 $.
In other words, the plasma was originally weakly non-ideal
($ {\Gamma}_e \sim 0.1 $), since its initial kinetic energy exceeded
the potential (Coulomb) energy only by one order of magnitude.

The overall plasma expansion rate was taken to be $ u_0^* = 0.1 $.
Then, ``physical'' size of the simulation cell (and, therefore, of the entire
plasma cloud) changes in the course of time according to
formula~(\ref{eq:length_unit_time}).
Since the simulations presented below in Figure~\ref{fig:Energy_time}
were conducted up to $ t^* \approx 600 $, the cloud expanded
approximately by 60~times and its density, respectively, dropped by
over 5~orders of magnitude.
Such variation well corresponds to the typical experimental
situations~\cite{Killian99,Killian07b}.

%%%%%%%%%%%%%%%%%%%%%%%%%%%%%%%%%%%%%
\begin{figure}
\includegraphics[width=1.0\columnwidth]{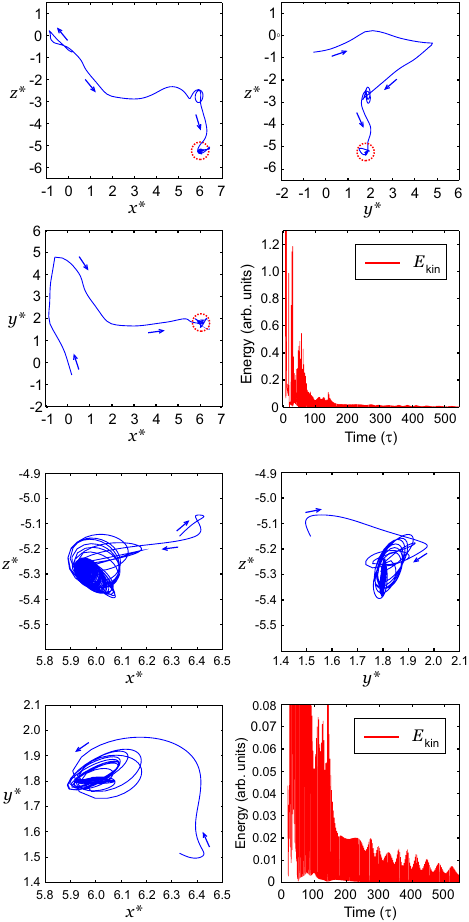}
\caption{\label{fig:Electr_shortper}
Trajectory of the first (short-period) captured electron viewed in three
coordinate planes at two different magnifications (three top and three bottom
panels, respectively) as well as its kinetic energy (the fourth panels in
the top and bottom).
The region of formation of a captured state of the electron is marked by
the red dotted circles.}
\end{figure}
%%%%%%%%%%%%%%%%%%%%%%%%%%%%%%%%%%%%%

%%%%%%%%%%%%%%%%%%%%%%%%%%%%%%%%%%%%%
\begin{figure}
\includegraphics[width=1.0\columnwidth]{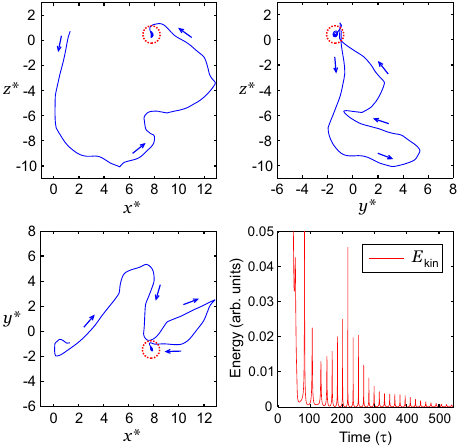}
\caption{\label{fig:Electr_longper}
Trajectory of the second (long-period) captured electron viewed in three
coordinate planes (two top and left bottom panels) as well as its kinetic
energy (right bottom panel).
The region of formation of a captured state of the electron is marked by
the red dotted circles.}
\end{figure}
%%%%%%%%%%%%%%%%%%%%%%%%%%%%%%%%%%%%%

Figure~\ref{fig:Energy_time} shows a temporal behavior of the total kinetic
and potential energies of all particles in the basic cell, including
the energy of interaction with the ``mirror'' images.
As should be expected, in the upper panel one can see initially a sharp decay
in the kinetic energy, which is just an adiabatic cooling of the expanding
plasma.
As a result, the electron Coulomb coupling parameter~$ {\Gamma}_e $ quickly
decays from~$ {\sim}0.1 $ to unity and approximately stabilizes at this
level.

The most interesting and nontrivial effect observed in this figure is
a development of fast oscillations (a series of the sharp equidistant
peaks), surviving up to the end of our simulation.
Moreover, starting from~$ t^* \approx 250{-}300 $ one can see onset of
the second, much longer period of oscillations, which also survives up
to the end of simulation.
A presence of the corresponding two periods is clearly seen in the second,
enlarged panel.
At last, inspection of the third, most magnified panel gives us a hint that
modulation of the kinetic and potential energies is caused just by
the passages of an electron, captured into a strongly elliptic orbit,
near the pericenter and apocenter of its trajectory.

A more careful analysis of the electron trajectories, presented in
Figs.~\ref{fig:Electr_shortper} and~\ref{fig:Electr_longper}, confirms
this conjecture.
It is seen that the above-mentioned electrons---after some intervals of
a ``random walk''---really form the confined states.
Their orbits are gradually contracting just because they are depicted
in the co-moving reference frame, expanding with the plasma cloud.
In fact, these pairs are detached from the overall expansion and have
an approximately constant size; but they look contracting in the scalable
coordinate system, where size of the basic cell~$ L^* $ remains constant.

The captured electrons should ultimately recombine with the corresponding
ions into the neutral atoms when quantum effects will be taken into account.
This corresponds to the well-known mechanism of the collisional two-stage
recombination~\cite{Massey56,Smirnov81}, whose total rate is governed
mostly by the first stage (\textit{i.e.}, formation of the classical bound
pairs).
The trajectories of other eight electrons in the basic cell do not exhibit
any features of localization.
Consequently, we can estimate the total efficiency of recombination as~20\%.
This is actually the main result of our simulations.

A further inspection of the trajectories shows that the first electron
(Fig.~\ref{fig:Electr_shortper}) became captured already at
$ t^* \approx 20{-}25  $ (\textit{i.e.}, when the plasma cloud expanded by
3--3.5~times); but its orbit initially was very irregular and, therefore,
did not exhibit a definite period.
Besides, it is interesting that the long-period modulation of energy,
caused by the second electron, is visible not only in the energy of the entire
system (Fig.~\ref{fig:Energy_time}) but also in the energy of the first
electron (bottom right panel in Fig.~\ref{fig:Electr_shortper}).
In general, this is not surprising, because the quickly-oscillating first
electron should naturally experience the long-period perturbations by
the second electron.

At last, inspection of the bottom right panel in Fig.~\ref{fig:Electr_longper}
shows that oscillations of energy of the second electron began to develop
already at $ t^* \approx 50{-}100 $, while they are visible in the total
energy (top panel in Fig.~\ref{fig:Energy_time}) only starting from
$ t^* \approx 250{-}300 $.
A gradual decay of the amplitude of these oscillations is evidently caused by
a decreasing ellipticity of the orbit.

% ====================
\section{Discussion}
\label{sec:Discussion}
% ====================

In the present work we followed the well-known idea of collisional
recombination~\cite{Massey56,Smirnov81} that the bound electron--ion
pairs are formed initially due to the classical dynamics, and then
almost all of them recombine to neutral atoms due to the quantum
transitions (which are commonly assumed to occur much faster).
So, the total efficiency of recombination should be equal approximately
to the rate of formation of the classically-bound electron--ion pairs.
(Of course, such an approach does not assume establishment of
the thermodynamic equilibrium.)

The plasma in our simulations was initially weakly-coupled
($ {\Gamma}_e \sim 0.1 $) but then quickly evolved to the moderately-coupled
state ($ {\Gamma}_e \sim 1 $), and the recombination processes developed
just in this regime.
Let us emphasize that the most of earlier UCP simulations were carried out
in a box of fixed size for a set of ``predefined'' parameters (density,
temperature, \textit{etc.}).
As distinct from that works, our method of ``scalable frames'' enables us
to simulate the expanding plasmas by a self-consistent way, \textit{i.e.},
the recombination occurs at the time-dependent density, temperature,
\textit{etc.}; and all these parameters are determined ``automatically''
from the same simulation.

Although the sample of particles employed in our work was rather small,
the results obtained turned out to be in agreement both with the experimental
measurements~\cite{Killian01} as well as with the earlier numerical
modeling~\cite{Niffenegger11}.
In fact, the criterion of recombination in the last-cited paper was
introduced rather arbitrarily as the percentage of electrons located closer
than 0.2~Wigner--Seitz radius to the nearest ion.
However, it is interesting that such choice of the ``critical'' radius
enabled that authors to get approximately the same total efficiency of
recombination (namely, $ 17{-}18\,\% $) as in our \textit{ab initio}
simulation.

Yet another frequently-used criterion~\cite{Lankin09b} is to find
the electrons that performed a specified minimal number of revolutions
(\textit{e.g.}, 1, 2, 4, or~6) about the nearest ion and to assume that
they will ultimately recombine to the neutral atoms.
In fact, this approach is even more popular and often considered as justified
better.
Unfortunately, analysis of the results of our simulations did not provide
a clear support in favor of such criterion.
So, this subject still needs to be studied in more detail.

As was already mentioned above, the employed numerical algorithm is
time-consuming: even at the relatively small number of particles in the basic
cell, each computational run on the ordinary PC (without parallelization)
took a few months.
So, when the number of particles was much greater, \textit{e.g.} 100 or
1000, we were able to perform simulations only for the considerably shorter
time intervals, below~$ 100\,\tau $, when the recombination processes were
not complete yet.
The low speed of calculations was caused by the fact that we used neither
the adaptive stepsize control for integration, nor Ewald summation technique
for the Coulomb sums.
However, one can expect that it will be possible in future to incorporate
these methods into our algorithm of the scalable (co-moving) reference
frames, which should make such simulations much faster.

In fact, it is not sufficiently clear how sensitive the recombination
statistics is to the accurate treatment of the long-range forces and if
the corresponding procedures can be reasonably simplified.
On the one hand, the bound states cannot be evidently formed by the two-body
interactions (since the total energy of the pair will be conserved in this
case).
So, the multi-particle interactions should be important in any
case~\cite{Massey56}.
On the other hand, one can try to describe them in a simpler approximation,
for example, by using the ``minimum-image convention'', \textit{i.e.},
summation over the particles located within the distance~$ \pm L/2 $ from
the given particle.
To the best of our knowledge, the earlier attempts by other authors
to employ the simplified summation schemes did not enable them to get
the sufficiently stable electron--ion pairs, but this subject still needs
to be studied in more detail.

Besides, it should be mentioned that a purely classical Coulomb system
is not stable in the genuine thermodynamic limit,
because the energy is unbounded from below and the trajectories can
collapse; so that the quantum-mechanical effects will be essential for
the stability of matter.
However, this argumentation is irrelevant to the case under consideration:
our system cannot reach the thermodynamic equilibrium because it steadily
expands up to infinity.
In such a case, the most evident stabilizing factor---preventing collapse
of the electron trajectories to the singularity---is an approximate
conservation (or, more exactly, adiabatic invariance) of the angular
momentum~$ M $ of a captured electron with respect to the corresponding
ion, which enters into the effective ``centrifugal''
potential~\cite{Landau76}:
\begin{equation}
U_{\rm eff} = \, \frac{M^2}{2 m_e r^2} \, - \, \frac{e^2}{r} \, .
\end{equation}
Then, the first term in the right-hand side prevents the electron orbit
from the collapse as long as $ M $~remains the adiabatic invariant.
In other words, the mechanism of (quasi-)stabilization in this case is
substantially different from any kinds of quantum corrections, such as
a finite electron thermal de~Broglie wavelength~\cite{Demyanov25},
\textit{etc.}

Let us emphasize that the numerical failures during ``head-on'' collisions,
which were discarded in our simulations, are absolutely irrelevant to
the above-mentioned collapse of the classical Coulomb systems in
the thermodynamic limit.
Really, the discarded runs typically involved the ``numerical pumping'' of
energy (\textit{i.e.}, ejection of the electrons) rather than a collapse
of their trajectories to the singularity of the potential.
In fact, we have already encountered the same numerical failures in our
previous works about the clusterized plasmas~\cite{Dumin22,Dumin24}.
This problem was easily resolved there by using the integration with
the adaptive stepsize control (ASSC), particularly, the subroutines from
the \textit{Numerical Recipes} library~\cite{Press92}.
Unfortunately, these subroutines become unreliable in the case of periodic
boundary conditions, which are essential for the present modeling.
So, as a simplest remedy, we just discarded the unfavorable initial
conditions.
In principle, it might be possible to modify the ASSC algorithms for
the periodic boundary conditions, but this will require considerable
efforts in the development of new software.

In fact, a lot of current works employ a much simpler approach, based
on the truncated Coulomb interactions; and they successfully reproduce
many properties of the ultracold plasmas.
However, we were afraid to do so in simulation of the recombination,
because the truncated Coulomb potential results in the inherently
unclosed classical orbits and, in principle, this can affect
the subsequent quantum transitions to the ground atomic states.

% ====================
\section{Conclusions}
\label{sec:Conclusions}
% ====================

\begin{enumerate}

\item
We suggested a new approach for modeling of the expanding ultracold
plasmas, which is based on the ``scalable'' reference frames, where
the effect of expansion formally manifests itself as an effective
viscous force in the co-moving coordinates.
This enabled us to simulate a non-stationary behavior of such plasmas
by a self-consistent way, when all physical parameters (density,
temperature, \textit{etc.}) are determined in the course of time by
the same model rather than specified in advance for a set of
simulations in the fixed-size boxes.

\item
To the best of our knowledge, the present work is the first successful
\textit{ab initio} simulation of recombination in the expanding ultracold
plasmas, which clearly demonstrates how some of electrons become localized
and form the stable compact pairs with the respective ions.
Such pairs are assumed to recombine subsequently to the ground atomic
states due to the quantum transitions.
It is important that we did not employ here any artificial criteria for
the formation of classical electron--ion pairs (such as a minimal number
of revolutions or a maximal distance between the particles).

\item
The efficiency of recombination at the realistic experimental conditions
obtained by our method turned out to be in agreement both with laboratory
measurements and the earlier semi-empirical estimates.
For example, if the initial value of the electron coupling
parameter~$ {\Gamma}_e $ was about~0.1 or somewhat less, and the plasma cloud
expanded by 50--60~times in radius (and, correspondingly, its density dropped
by over 5~orders of magnitude), about 20\% of the original electrons
recombined, which coincides perfectly with the experiment~\cite{Killian01}.
The earlier numerical modeling~\cite{Niffenegger11} resulted in a similar
value, 17--18\%; but this coincidence should be taken with caution, because
the corresponding simulation was carried out in a box of fixed size, and
the recombined particles were identified by a semi-empirical criterion.

\item
Unfortunately, the proposed modeling scheme is rather expensive
computationally: for example, a single run takes a few months on
the ordinary PC.
However, we expect that its efficiency can be improved in future with
an optimized software, which should employ the Ewald summation technique
for the Coulomb forces and the adaptive stepsize control for integration.

\end{enumerate}

\begin{acknowledgments}

YVD is grateful to
A.A.~Bobrov,
P.R.~Levashov,
S.A.~Mayorov,
J.-M.~Rost,
S.A.~Saakyan,
U.~Saalmann,
V.S.~Vorob'ev,
B.B.~Zelener, and
B.V.~Zelener
for fruitful discussions and valuable suggestions as well as to
O.E.~Stakheev for the assistance in numerical calculations.
We are also grateful to
O.~Primina and
A.~Truskov
for the support and encouragement of this work.

\end{acknowledgments}

\section*{Author declarations}

\subsection*{Conflict of interest}

The authors have no conflicts to disclose.

% \subsection*{Author contributions}

\section*{Data availability}

The data that support the findings of this study are available from the
corresponding author upon reasonable request.

\appendix

\section{Examples of the ``favorable'' and ``unfavorable''
         initial conditions}
\label{sec:Examples}

%%%%%%%%%%%%%%%%%%%%%%%%%%%%%%%%%%%%%
\begin{figure}
\includegraphics[width=0.9\columnwidth]{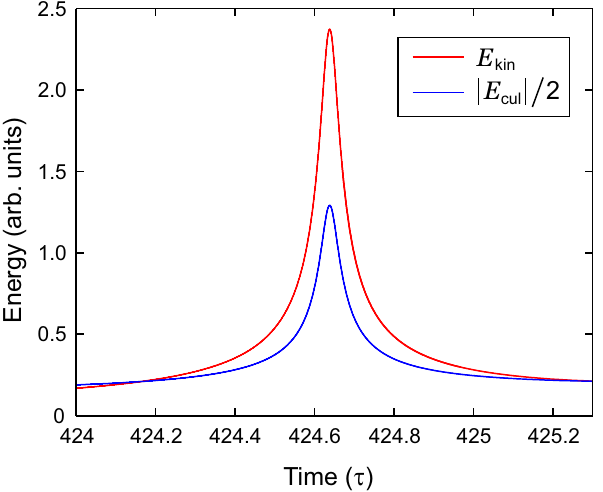}
\caption{\label{fig:Ex_resolved}
Temporal variations of the kinetic and potential energies of all particles
in the basic cell in the case of ``favorable'' initial conditions (which are
the same as in Figs.~\ref{fig:Energy_time}--\ref{fig:Electr_longper}).}
\end{figure}
%%%%%%%%%%%%%%%%%%%%%%%%%%%%%%%%%%%%%

%%%%%%%%%%%%%%%%%%%%%%%%%%%%%%%%%%%%%
\begin{figure}
\includegraphics[width=0.9\columnwidth]{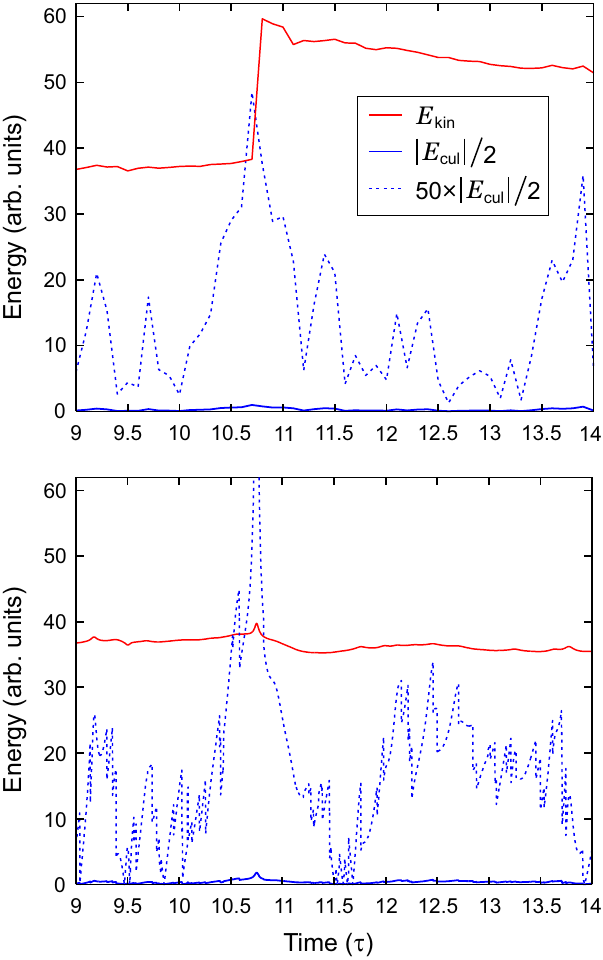}
\caption{\label{fig:Ex_unresolved}
Temporal variations of the kinetic and potential energies in the case of
``unfavorable'' initial conditions when integration was performed with
the standard (top panel) and enhanced accuracy (bottom panel).}
\end{figure}
%%%%%%%%%%%%%%%%%%%%%%%%%%%%%%%%%%%%%

As was already mentioned in Sec.~\ref{sec:Integr_method}, the employed
integrator was sometimes unable to resolve reliably the close passages
of electrons near the ions in the very beginning of simulations, and
the respective initial conditions were discarded.
Here we provide the corresponding examples.

Figure~\ref{fig:Ex_resolved} represents the case of a well-resolved passage
of an electron near the pericenter of its strongly-elliptic orbit at
the sufficiently large time, when the stable electron--ion pairs were
already formed.
(The initial conditions are the same as in
Figs.~\ref{fig:Energy_time}--\ref{fig:Electr_longper}.)
One can see here two major features in variations of the kinetic and
potential energies:
Firstly, the energetic peaks are symmetric, as should be expected for
motion in the elliptic orbit (if weak perturbations from the remote
particles are ignored).
Secondly, the peaks in kinetic and absolute value of potential energies
are approximately of the same magnitude, as should be naturally
expected from conservation of the total energy.
Really, the potential (Coulomb) energy is actually negative and,
in the case of strongly-elliptic orbit, the total energy of the pair
is much less than the above-mentioned peak values (\textit{i.e.},
roughly speaking, is close to zero).

An absolutely different situation can be seen in
Fig.~\ref{fig:Ex_unresolved}, which corresponds to the case of
``unfavorable'' initial conditions.
In general, a very chaotic pattern of variations is not surprising,
because this is an early stage of the plasma cloud expansion, when
relaxation did not complete yet and the plasma remained rather hot.
One can see in the top panel a sharp asymmetric jump of the kinetic
energy at~$ t^* = 10.7 $, which is evidently unphysical and caused
just by numerical errors.
Really, this increase in the kinetic energy (about 20~arbitrary units)
is an order of magnitude greater than the associated increase in
the absolute value of potential energy (about 2~units), which violates
the energy conservation condition.
Next, the peak of potential energy is approximately symmetric
(because position of the electron always changes smoothly with
respect to the ion), but the kinetic energy exhibits a strongly
asymmetric jump (step).
In fact, such cases are well known in UCP simulations and commonly
called the ``numerical heating''.

However, when the accuracy of numerical integration is improved
(bottom panel in Fig.~\ref{fig:Ex_unresolved}),
the above-mentioned jump transforms into a tiny symmetric peak,
whose amplitude is comparable to the peak of potential energy,
as should be expected for a close interparticle collision.
Therefore, the problem was purely numerical.
Unfortunately, because of the limited computational resources, we
were unable to perform the entire set of simulations with
the substantially improved accuracy.
So, the ``unfavorable'' cases were just disregarded.

\subsection*{References:}


% \bibliography{Dumin}

\end{document}